\documentclass[twocolumn]{emulateapj}

\newcommand{\Msun}{${\rm M}_\odot$}
\newcommand{\Msunyr}{${\rm M}_\odot$yr$^{-1}$}
\newcommand{\Msunyrkpcsq}{${\rm M}_\odot$yr$^{-1}$kpc$^{-2}$}
\newcommand{\SFRSD}{$\Sigma_{\rm SFR}$} 
\newcommand{\DustSD}{$\Sigma_{\rm dust}$} 
\newcommand{\SigmaMstar}{$\Sigma_{\rm M*}$} 

\newcommand{\uJyperbm}{$\mu$Jy\,beam$^{-1}$} 

\begin{document}


\title{VLA and ALMA Imaging of Intense, Galaxy-Wide Star Formation in $\lowercase{\it z} \sim$ 2 Galaxies}


\author{W. Rujopakarn\altaffilmark{1,2}, 
J. S. Dunlop\altaffilmark{3},
G. H. Rieke\altaffilmark{4},
R. J. Ivison\altaffilmark{3,5},
A. Cibinel\altaffilmark{6},
K. Nyland\altaffilmark{7},
P. Jagannathan\altaffilmark{8},\\
J. D. Silverman\altaffilmark{1},
D. M. Alexander\altaffilmark{9},
A. D. Biggs\altaffilmark{5},
S. Bhatnagar\altaffilmark{8},
D. R. Ballantyne\altaffilmark{10},
M. Dickinson\altaffilmark{11},\\
D. Elbaz\altaffilmark{12},
J. E. Geach\altaffilmark{13},
C. C. Hayward\altaffilmark{14,15},
A. Kirkpatrick\altaffilmark{16},
R. J. McLure\altaffilmark{3},
M. J. Micha{\l}owski\altaffilmark{3},\\
N. A. Miller\altaffilmark{17},
D. Narayanan\altaffilmark{18},
F. N. Owen\altaffilmark{8},
M. Pannella\altaffilmark{19},
C. Papovich\altaffilmark{20},
A. Pope\altaffilmark{21},
U. Rau\altaffilmark{8},\\
B. E. Robertson\altaffilmark{22},
D. Scott\altaffilmark{23},
A. M. Swinbank\altaffilmark{9},
P. van der Werf\altaffilmark{24},
E. van Kampen\altaffilmark{5}, 
B. J. Weiner\altaffilmark{4}, \\  and 
R. A. Windhorst\altaffilmark{25}
}

\affil{$^1$Kavli Institute for the Physics and Mathematics of the Universe (WPI), The University of Tokyo Institutes for Advanced Study, \\ The University of Tokyo, Kashiwa, Chiba 277-8583, Japan; wiphu.rujopakarn@ipmu.jp\\
$^2$Department of Physics, Faculty of Science, Chulalongkorn University, 254 Phayathai Road, Pathumwan, Bangkok 10330, Thailand\\
$^3$Institute for Astronomy, University of Edinburgh, Royal Observatory, Blackford Hill, Edinburgh EH9 3HJ, UK\\
$^4$Steward Observatory, University of Arizona, Tucson, AZ 85721, USA\\
$^5$European Southern Observatory, Karl Schwarzschild Strasse 2, Garching, Germany\\
$^6$Astronomy Centre, Department of Physics and Astronomy, University of Sussex, Brighton, BN1 9QH, UK\\
$^7$National Radio Astronomy Observatory, Charlottesville, VA 22903, USA\\
$^8$National Radio Astronomy Observatory, Socorro, NM 87801, USA\\
$^9$Department of Physics, Durham University, Durham DH1 3LE, UK\\
$^{10}$Center for Relativistic Astrophysics, School of Physics, Georgia Institute of Technology, Atlanta, GA 30332, USA\\
$^{11}$National Optical Astronomy Observatory, 950 North Cherry Avenue, Tucson, AZ 85719, USA\\
$^{12}$CEA Saclay, DSM/Irfu/Service d'Astrophysique, Orme des Merisiers, F-91191 Gif-sur-Yvette Cedex, France\\
$^{13}$Center for Astrophysics Research, Science \& Technology Research Institute, University of Hertfordshire, Hatfield AL10 9AB, UK\\
$^{14}$TAPIR 350-17, California Institute of Technology, 1200 E. California Boulevard, Pasadena, CA 91125, USA\\
$^{15}$Harvard-Smithsonian Center for Astrophysics, 60 Garden Street, Cambridge, MA 02138, USA\\
$^{16}$Yale Center for Astronomy \& Astrophysics, Physics Department, P.O. Box 208120, New Haven, CT 06520, USA\\
$^{17}$Department of Mathematics and Physical Sciences, Stevenson University, Stevenson, MD 21153-0641, USA\\
$^{18}$Department of Astronomy, University of Florida, Gainesville, FL 32611-2055, USA\\
$^{19}$Faculty of Physics, Ludwig-Maximilians Universit\"at, Scheinerstr. 1, 81679 Munich, Germany\\
$^{20}$Department of Physics and Astronomy, Texas A\&M University, College Station, TX, 77843-4242 USA\\
$^{21}$Department of Astronomy, University of Massachusetts, Amherst, MA 01003, USA\\
$^{22}$Department of Astronomy and Astrophysics, University of California, Santa Cruz, CA 95064, USA\\
$^{23}$Department of Physics and Astronomy, University of British Columbia, 6224 Agricultural Road, Vancouver, BC, V6T1Z1 Canada\\
$^{24}$Leiden Observatory, Leiden University, PO Box 9513, 2300 RA Leiden, The Netherlands\\
$^{25}$School of Earth and Space Exploration, Arizona State University, Tempe, AZ 85287, USA
}




\begin{abstract}

We present $\simeq0\farcs4$-resolution extinction-independent distributions of star formation and dust in 11 star-forming galaxies (SFGs) at $z = 1.3-3.0$. These galaxies are selected from sensitive, blank-field surveys of the $2' \times 2'$ Hubble Ultra-Deep Field at $\lambda = 5$ cm and 1.3 mm using the Karl G. Jansky Very Large Array (VLA) and Atacama Large Millimeter/submillimeter Array (ALMA). They have star-formation rates (SFRs), stellar masses, and dust properties representative of massive main-sequence SFGs at $z \sim 2$. Morphological classification performed on spatially-resolved stellar mass maps indicates a mixture of disk and morphologically disturbed systems; half of the sample harbor X-ray active galactic nuclei (AGN), thereby representing a diversity of $z \sim 2$ SFGs undergoing vigorous mass assembly. We find that their intense star formation most frequently occurs at the location of stellar-mass concentration and extends over an area comparable to their stellar-mass distribution, with a median diameter of $4.2 \pm 1.8$ kpc. This provides direct evidence for galaxy-wide star formation in distant, blank-field-selected main-sequence SFGs. The typical galactic-average SFR surface density is 2.5 M$_{\sun}$yr$^{-1}$kpc$^{-2}$, sufficiently high to drive outflows. In X-ray-selected AGN where radio emission is enhanced over the level associated with star formation, the radio excess pinpoints the AGN, which are found to be co-spatial with star formation. The median extinction-independent size of main-sequence SFGs is two times larger than those of bright submillimeter galaxies whose SFRs are $3-8$ times larger, providing a constraint on the characteristic SFR ($\sim300$ M$_{\sun}$yr$^{-1}$) above which a significant population of more compact star-forming galaxies appears to emerge.

\end{abstract}





\section{Introduction} \label{sec:intro}

Numerical simulations and observational inferences suggest that typical star-forming galaxies (SFGs) at the peak of galaxy assembly activity, $z \simeq 1 - 3$, assembled most of their stellar mass via accretion of cold gas, which led to gas-rich, unstable disks and disk-wide star formation \citep[e.g.,][]{Keres05, Dekel09, Bournaud09}. The isolated, in-situ assembly of typical SFGs is inferred from the presence of a relationship between star formation and stellar mass, the ``main sequence'' \citep[e.g.,][]{Noeske07, Whitaker12, Speagle14}, and from the rarity of compact starbursts at $z \sim 2$, as indicated by their specific star-formation rate (SFR) distribution and infrared color \citep{Rodighiero11, Elbaz11}. At higher SFRs ($\gtrsim300$ \Msunyr\ at $z \sim 2$), there is no theoretical consensus on whether mergers or continuous accretion are the dominant triggering mechanism of intense star formation \citep{Hopkins10, Dave10, Hayward13, Narayanan15}; the relative contribution of in-situ versus merger modes at this SFR regime have not been constrained observationally. 

Contrary to the inferred in-situ assembly of SFGs, {\it Hubble Space Telescope} ({\it HST}) observations of main-sequence SFGs at $z \sim 2$ commonly show galactic substructures and asymmetry that are signposts of mergers that drive intense star formation in the local Universe \citep{Lotz04, Kartaltepe12}. However, some substructures, such as optically-bright star-forming clumps with SFR $\simeq1 - 30$  of \Msunyr\ \citep[e.g.,][]{FS09, Guo15} can also be a natural consequence of gas-rich, turbulent disks evolving in isolation. Yet the role of star-forming clumps in assembling the bulk of stellar mass is debated \citep{Genel12, Bournaud14} and optically-selected clumps altogether contain $< 10-20\%$ of the total star formation of their host \citep{Wuyts12, Guo12, Guo15}. Directly observing the distribution of the bulk of star formation in galaxies at $z \sim 2$ is therefore key to establishing the relative contribution between modes of star and bulge formation.

Significant progress in directly imaging the distribution of star formation at $z > 1$ has been made with spatially-resolved H$\alpha$ spectroscopy, e.g., SINS \citep{FS09}, KMOS$^{\rm 3D}$ \citep{Wisnioski15}, and KROSS \citep{Stott16}. Average H$\alpha$ maps from 3D-HST, the largest sample thus far of resolved star formation at $z = 1.5 - 2.5$, show that the star-formation surface density, \SFRSD, on average peaks near the centers of massive galaxies \citep{Nelson15}. However, the dust extinction also peaks at the center, such that a factor of 6 correction to the inferred H$\alpha$ SFR is required in the central kpc \citep[and more than a factor of 10 close to the center;][]{Nelson16}. Above SFRs as small as 20 \Msunyr, which is $0.2\times$ the typical rate for main-sequence SFGs at $z > 1$, galaxies become so dust enshrouded that almost no light emerges in rest-frame ultraviolet observations \citep{Reddy10}. The question of where exactly new stars form within typical $z \sim 2$ SFGs is hence a deceptively simple one that is challenging to address.

Breakthroughs in this area require sub-arcsecond resolution, extinction-independent star-formation tracers for main-sequence SFGs at $z > 1$, which are now available with the Karl G. Jansky Very Large Array (VLA) and the Atacama Large Millimeter/submillimeter Array (ALMA). To this end, we conduct two sensitive, blank-field imaging surveys of the Hubble Ultra-Deep Field (HUDF, $\alpha = 03^{\rm h}$32$^{\rm m}$, $\delta =-27^{\circ}47'$) using the VLA and ALMA at $\lambda = 5$ cm and 1.3 mm, respectively, to make $0\farcs4$ resolution images of SFGs at $z \sim 2$. The 5 cm continuum traces star formation through the synchrotron emission from supernova remnants, but can be affected by AGN emission; whereas the 1.3 mm continuum traces cold dust associated with star formation, but requires uncertain assumptions about the shapes of spectral energy distributions (SEDs) to estimate the SFR. The combination of the two surveys therefore provides complementary strengths, especially in the HUDF where the wealth of ancillary data can help, e.g., identify AGN. By establishing that the VLA and ALMA trace the common extent of star formation, ALMA can serve as a morphological tracer of ``pure'' star formation in AGN hosts because the 1.3-mm dust continuum is neither contaminated by AGN torus emission \citep{Elvis94, Mullaney11, MorNetzer12} nor synchrotron emission from the jets\footnote{Assuming a flat spectral index for AGN, the radio continuum from the brightest radio AGN in the HUDF is less than 1\% of their 1.3 mm emission from star formation.}. Furthermore, because ALMA significantly gains in sensitivity to star formation with the help of negative K-corrections at $z > 2.5$, whereas the VLA gradually loses star-formation sensitivity beyond this redshift, the combination of VLA and ALMA yields a sensitive probe of the morphology of star formation over the entire range of $z = 1 - 3$. 

In this paper, we present first results from combining the VLA and ALMA HUDF surveys, focusing on the extinction-independent distributions of the star formation in SFGs at $z = 1-3$. We discuss the VLA and ALMA surveys and ancillary data in \S\ref{sec:observations} and present the size and location of star-formation in SFGs selected from VLA and ALMA, along with their implications in \S\ref{sec:results}. We adopt a $\Lambda$CDM cosmology with $\Omega_M = 0.3$, $\Omega_\Lambda = 0.7$, $H_{0} = 70~{\rm km\,s}^{-1}{\rm Mpc}^{-1}$, and the \citet{Chabrier03} IMF. 

\begin{figure*}[t]
\figurenum{1}
\centerline{\includegraphics[width=0.85\textwidth]{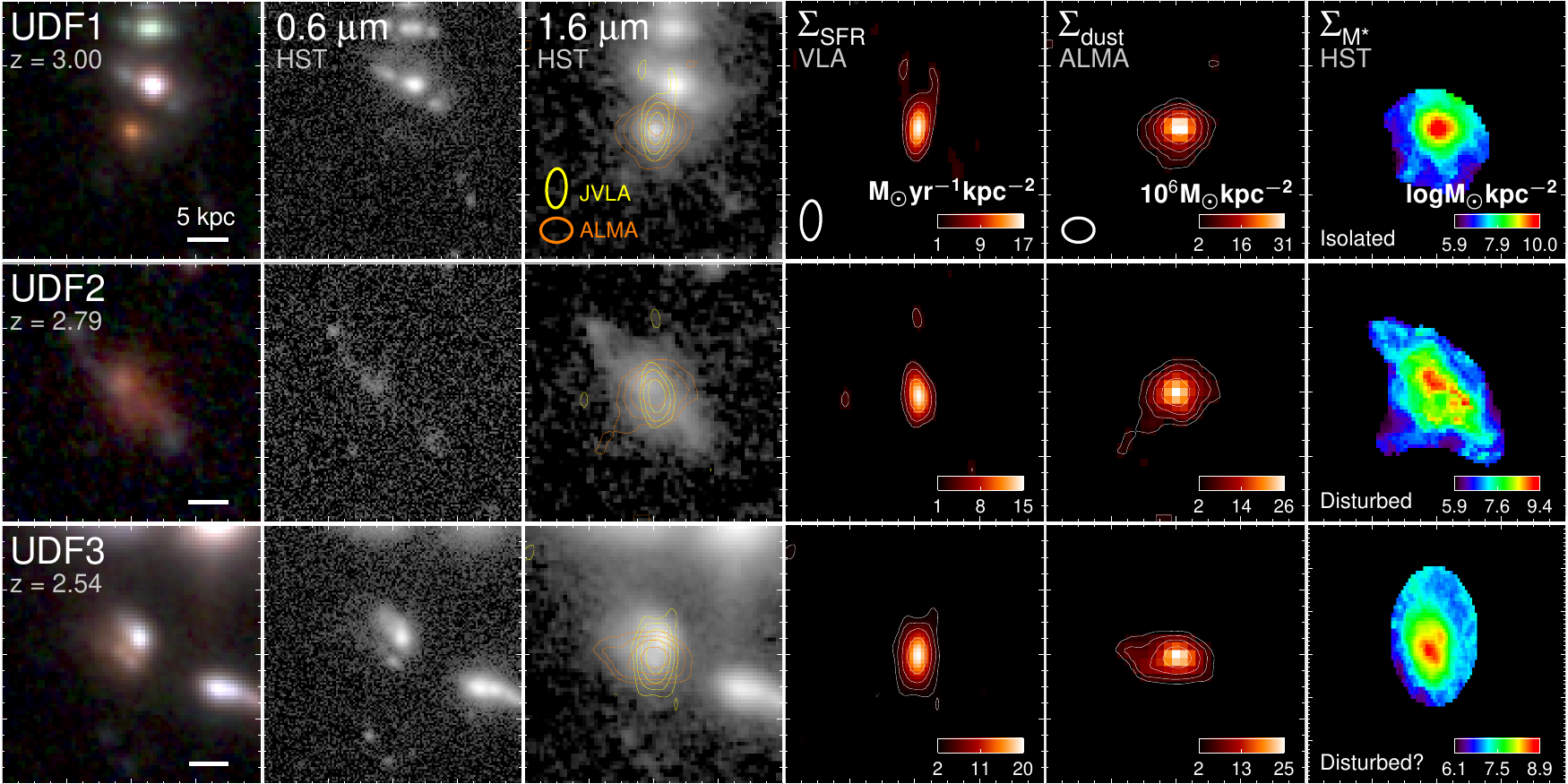}}
\caption{The 11 SFGs detected in both the VLA and ALMA HUDF surveys. From left to right: {\it HST} $i_{814}-J_{125}-H_{160}$ color composite; unobscured star formation ({\it HST}/F606W); rest-frame optical morphology ({\it HST}/F160W); star-formation rate surface density, \SFRSD\ (VLA, details in \S\ref{sec:observations}); dust mass surface density, \DustSD\ (ALMA, details in \S\ref{sec:gas}); and stellar-mass surface density, \SigmaMstar\ (from spatially-resolved SED fitting using {\it HST} images). Each image is $4'' \times 4''$; North is up, East is on the left. VLA and ALMA synthesized beams are shown in the corresponding columns; the contours are $[-3, 3^{1}, 3^{1.5}, 3^{2}, ...] \times \sigma$ for VLA and $[-2.5, 2.5^{1}, 2.5^{1.5}, 2.5^{2}, ...] \times \sigma$ for ALMA; negative contours are shown as dotted lines. Intense star formation most frequently occurs within the stellar-mass concentration and extends over a large area of the stellar-mass buildup, i.e., galactic-wide (continued on next page). \label{fig:sixpanA}}
\end{figure*}

\begin{figure*}[t]
\figurenum{1}
\centerline{\includegraphics[width=0.85\textwidth]{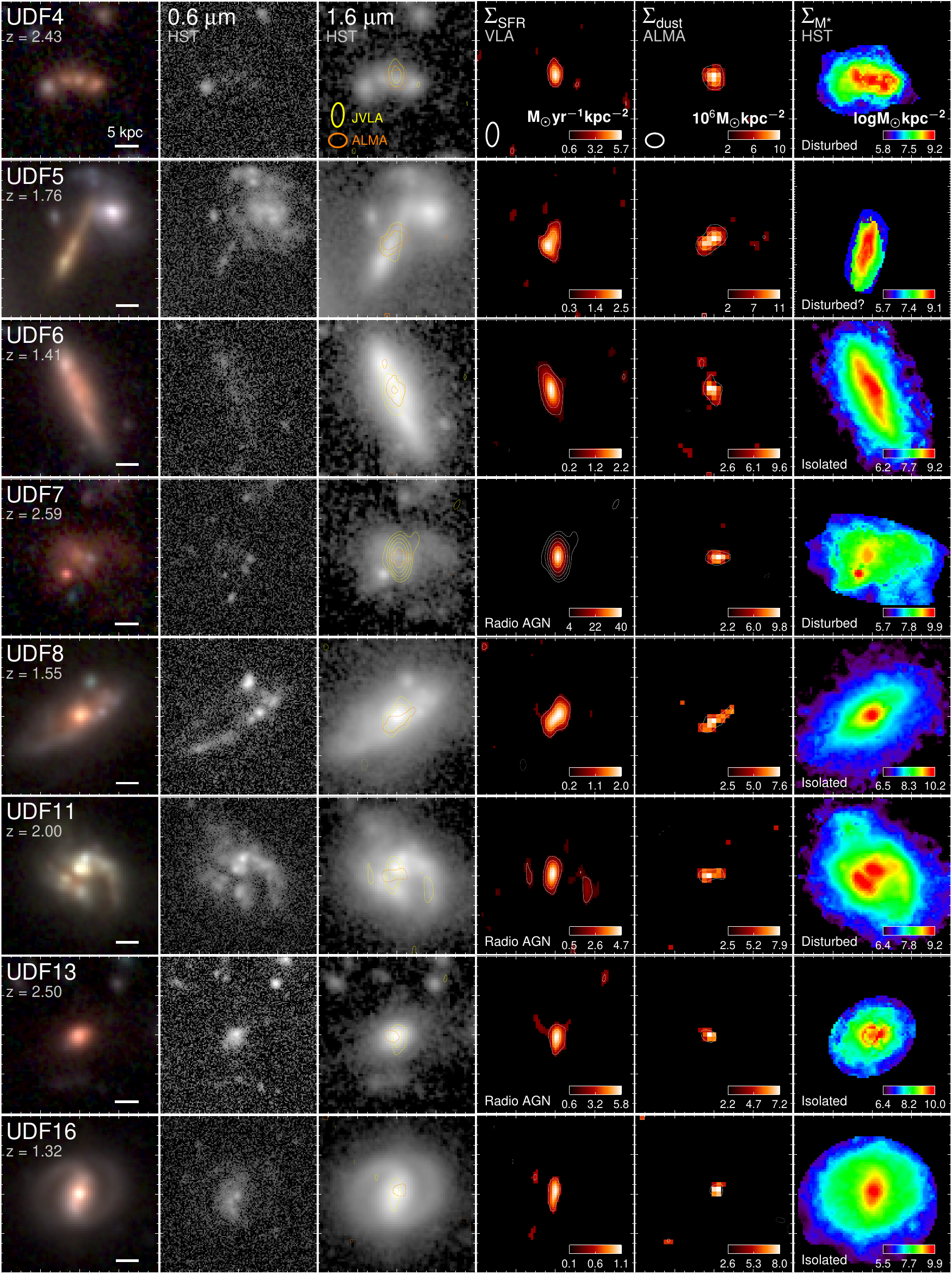}}
\caption{(continued) We note that the radio \SFRSD\ maps for UDF7, 11, and 13 that harbor X-ray-selected AGN with radio emission enhanced more than twice the level of star-formation emission (Figure \ref{fig:FRC}, tabulated in Table \ref{tab:sourcetable}) may contain significant contributions from the AGN and hence marked as ``radio AGN''.}
\end{figure*}

\section{Observations and Ancillary Data} \label{sec:observations}

To image SFGs at $z \sim 2$ with the VLA, the radio continuum at 5\,cm offers a good balance between angular resolution and the expected flux density, given the synchrotron radio spectrum, S$_{\nu} \propto \nu^{-0.7}$. We observed the HUDF for 177 hours in the A, B, and C configurations during 2014 March $-$ 2015 September using the full C-band bandwidth of $4 - 8$ GHz ($\lambda = 7.5-3.7$\,cm). The observations comprised 42 dynamically-scheduled sessions of $2.5 - 5.5$\,hours. Each session observed 3C48 for flux and bandpass calibrations; J0402-3147 was observed for phase calibration every 25 minutes. Data reduction was carried out with CASA \citep{McMullin07} using the following steps: (1) standard calibration using the VLA Data Reduction Pipeline (Chandler et al., in prep); (2) removal of any portions of the data corrupted by strong radio frequency interference; and (3) imaging with the task {\tt TCLEAN}. The imaging parameters are the following: MT-MFS deconvolver with nterms of 2, $0\farcs06$ pixel size, and Briggs weighting with robust parameter of 0.5. We imaged the data well beyond the primary beam radius of $3.6'$ (employing the w-projection) to mitigate the imaging artifacts caused by the sidelobes from bright sources far from the pointing center. The final image has a $0\farcs31 \times 0\farcs61$ synthesized beam and rms noise at the pointing center of 0.32 \uJyperbm, consistent with the theoretical sensitivity (0.30 \uJyperbm).

The ALMA HUDF Deep Field is a 1.3-mm survey of the 4.5-arcmin$^2$ HUDF using a 45-pointing mosaic during 2014 July $-$ 2015 May in 13 sessions, utilizing a total of approximately 20\,hours. The rms noise of the naturally-weighted ALMA map is 29 \uJyperbm\ and the synthesized beam is $0\farcs37 \times 0\farcs48$. Each session observed J0334-301 for flux and bandpass calibrations, and also for phase calibration during the 2015 sessions (2014 sessions used J0348-2749 for phase calibration). Calibration was carried out with CASA and imaging with the task {\tt CLEAN}, adopting natural weighting to maximize sensitivity. Deconvolution was not performed as there are no strong sidelobes from 1.3-mm objects in the HUDF. Details of the ALMA observations, data reduction, and source extraction are given in a companion paper on the ALMA HUDF survey \citep[hereafter, D16]{Dunlop16}. 

The following physical parameters in this paper are also estimated by D16: (1) SFR$_{\rm IR}$ that combines {\it Spitzer} (24 \micron), {\it Herschel} (deblended $70 - 500$ \micron), and ALMA measurements; (2) galaxy-integrated stellar mass via SED fitting; (3) spectroscopic redshift compilation, using redshift measurements, e.g., from VLT/MUSE IFU survey (Bacon et al., in preparation), and photometric redshifts where spectroscopic redshift are not available. We identify X-ray AGN using the 4 Ms {\it Chandra} catalog \citep{Xue11} with updated redshifts and estimate the 5 cm radio SFR using the \citet{Bell03} indicator, assuming a spectral slope S$_{\nu} \propto \nu^{-0.7}$ for K-correction.

{\it HST} images of the HUDF reach $29.5 - 30.3$ mag (AB) at $0.4 - 1.6$ \micron\ \citep{Ellis13, Koekemoer13}, from which we construct stellar-mass maps by fitting spatially-resolved SEDs using the procedures described by \citet{Cibinel15}. The sums of the stellar mass in these maps agree within 0.1 dex with the integrated stellar-mass estimates that utilize longer wavelength photometry \citep[e.g., {\it Spitzer}/IRAC $3.6 - 8.0$ \micron; ][]{Cibinel15}, suggesting that the maps provide a good description of stellar-mass distribution in obscured SFGs. These stellar-mass maps serve as a reference frame to map where star formation is occurring in relation to stellar-mass buildup within each galaxy. We further classify the stellar-mass morphologies as isolated or disturbed by performing the asymmetry \citep{Conselice03, Zamojski07} and $M_{20}$ \citep{Lotz04} analysis on these stellar-mass maps, with the $M_{20}$ being the second-order moment of the 20\% brightest pixels, following \citet{Lotz04}. This classification method is shown by \citet{Cibinel15} to be capable of identifying galaxy mergers with a smaller fraction of contamination from clumpy disks compared to single-band classifications. Specifically, \citet{Cibinel15} demonstrated that the asymmetry $-$ $M_{20}$ classification results in $\lesssim 20\%$ contaminations when performed on stellar-mass maps, whereas the contamination fraction is $\sim 50\%$ using the single-band {\it HST}/F160W images alone. This is because rest-frame optical images often contain optically-bright clumps of star formation in addition to the dominant stellar-mass concentrations; spatially-resolved SED fitting is required to distinguish them.

The positional accuracy, $\sigma_{\rm pos}$, with which we can pinpoint the locations of star formation from the VLA and ALMA images depends on: (1) the positional accuracy of the phase calibrators used as the astrometric reference (these errors are $< 2$ mas for our observations); and (2) the SNR of the detection, since $\sigma_{\rm pos} \approx \theta_{\rm beam}$/(2~SNR), following \citet{Condon97}, corresponding to $\simeq$ 40 and 60 mas for VLA and ALMA, respectively, at their detection limits. Comparing optical, radio, and millimeter morphologies further requires accurate astrometric alignment between the wave bands. The positions of emission peaks from the VLA and ALMA images are mutually consistent within 40 mas, corresponding to 0.3 kpc at $z = 2$, indicating good astrometric agreement despite observing different phase calibrators, and neither shows systematic offsets compared to 2MASS positions. The VLA primary beam extends beyond the HUDF to a total area of 61 arcmin$^2$ with rms sensitivity better than 1 \uJyperbm, co-spatial with the CANDELS imaging \citep{Grogin11, Koekemoer11}. This area contains 68 bright point sources ($>8\sigma$) detected in both VLA and {\it HST}/F160W images that we use to compare VLA's astrometry (and by proxy, ALMA's) against {\it HST}'s. An offset of $\Delta\alpha = -80 \pm 110$ mas, $\Delta\delta = 260 \pm 130$ mas is required to bring {\it HST} astrometry into agreement with those of VLA and ALMA. This offset is constant throughout the field (a possible source of offset is discussed in D16). We apply the offset to all {\it HST} images for further analysis; the resulting median systematic offset is $<10$ mas in both $\alpha$ and $\delta$, with an rms dispersion of 150 mas.

\begin{figure*}
\figurenum{2}
\centerline{\includegraphics[width=0.88\textwidth]{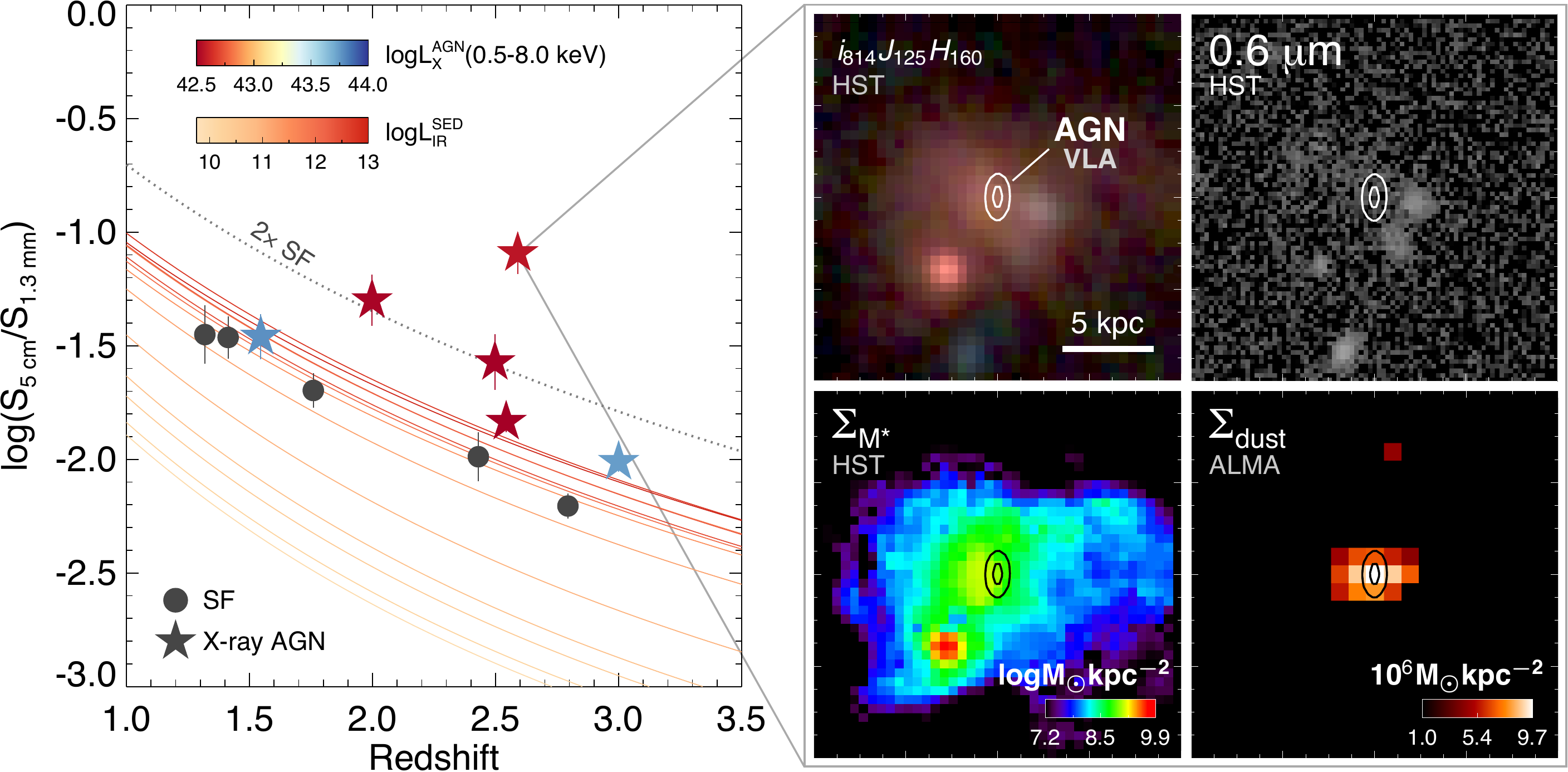}}
\caption{Left: Ratios of radio and millimeter fluxes for SFGs are well described by the spectral energy distribution (SED) of local galaxies with infrared luminosity of $10^{11.5}$ L$_{\odot}$ (flux ratios predicted by the \citet{Rieke09} library shown in red curves, color-coded by the SED template's infrared luminosity; the ratio twice the maximum level for SFGs is shown by the dotted line), confirming that the far-infrared/radio correlation holds for individual galaxies down to the level of typical SFGs out to $z \sim 3$. On the other hand, some X-ray selected AGN, particularly the X-ray fainter ones, have strongly enhanced radio emission (AGN are plotted as stars color-coded by their X-ray luminosities at $0.5-8$ keV). Right: VLA pinpoints the location of the AGN in UDF7 whose radio emission is AGN-dominated. Clockwise from top-left are {\it HST} $i_{814}-J_{125}-H_{160}$ color composite, rest-frame UV image, \DustSD, and \SigmaMstar\ maps; each cutout is $2\farcs4 \times 2\farcs4$; the contours mark the radio detection at 35 and 40$\sigma_{\rm VLA}$. In this example, both the AGN and dust associated with star formation are co-spatial, but occur in a location with low stellar-mass surface density; neither are visible in the rest-frame UV image. \label{fig:FRC}}
\end{figure*}

\section{Results} \label{sec:results}

Our sample is based on the ALMA-selected SFGs at $z = 1-3$, which D16 has demonstrated are representative of massive main-sequence SFGs at these redshifts. At $z = 1-3$, the D16 sample contains 13 galaxies detected in the ALMA image at $\geqslant3.5\sigma$, 11 of these are detected at $5 - 30\sigma$ in the VLA image; the remaining two, UDF10 and 15 (see Table 2 of D16), are detected at $2.5-3.0\sigma$ in the VLA image (they are $3.6-4.0\sigma$ in the ALMA image), which we exclude from further morphological analysis. Since their stellar masses and rest-frame optical extents (an upper limit of star-formation size) are similar to the rest of the sample, their exclusion will not bias our conclusions. The final sample contains 11 galaxies selected from ALMA and VLA, tabulated in Table 1 and shown in Figure \ref{fig:sixpanA}. The sample has a median redshift of $z = 2.2$ and SFRs and stellar masses ranging from $40-326$ \Msunyr\ (median $= 102$ \Msunyr) and $10.3 \leqslant$ log(M$_*$/\Msun) $ \leqslant 11.2$ (median $= 10.7$), respectively, implying vigorous assembly of stellar mass with a median mass-doubling time of 0.4 Gyr. It is worth noting that there are six VLA-detected galaxies in the HUDF in the same redshift range that are not detected by ALMA, these galaxies are at $z_{\rm median} = 1.3$ where VLA is sensitive to lower SFRs than ALMA (e.g., at $z = 1.0$, the VLA detection limit is $\approx 10$ \Msunyr, whereas that of ALMA is $\approx60$ \Msunyr). Likewise, their exclusion will not bias the conclusion for massive galaxies above $10^{10}$ \Msun. The morphological classification performed on stellar-mass maps (\S\ref{sec:observations}) indicates that five and four galaxies are unambiguously isolated and morphologically disturbed, respectively; the remaining three exhibit sub-structures in stellar-mass maps that could indicate, e.g., late-stage mergers (classifications indicated in Figure \ref{fig:sixpanA}).

We will first show that individual SFGs follow the far-infrared/radio correlation (\S\ref{sec:FRC}), then report on the sizes and locations of their star formation in relation to the stellar-mass buildup of host galaxies (\S\ref{sec:locandsizes}), the dust and gas masses and the implied gas fraction and depletion time (\S\ref{sec:gas}), the spatially-resolved \SFRSD\ and their potential implications on star formation-driven outflows (\S\ref{sec:sigmasfr}), as well as discussing our general observations on star formation in AGN hosts (\S\ref{sec:SFinAGN}).

\begin{deluxetable*}{lcccccccccc}
\tablenum{1}
\tablecaption{VLA and ALMA-selected of Star-Forming Galaxies at $z = 1-3$ in the HUDF\label{tab:sourcetable}}
\tablehead{
\colhead{ID} & \colhead{$\alpha_{\rm radio}$} & \colhead{$\delta_{\rm radio}$} & \colhead{$z$} & \colhead{M$_*$} & \colhead{AGN?} & \colhead{SFR$_{\rm radio}$} & \colhead{SFR$_{\rm IR}$} & \colhead{M$_{\rm dust}$} & \colhead{$f_{\rm gas}$} & \colhead{$\tau_{\rm dep}$} \\
\colhead{} & \colhead{(deg)} & \colhead{(deg)} & \colhead{} & \colhead{(log\Msun)} & \colhead{} & \colhead{(\Msunyr)} & \colhead{(\Msunyr)} & \colhead{(log\Msun)} & \colhead{} & \colhead{(Gyr)}
}
\startdata
UDF1   & 53.18346 & $-$27.77664 & 3.00\phantom{1}  & $10.7 \pm 0.1$ & X-ray & $<476      $ & $326 \pm  83$     & $ 9.1^{+0.2}_{-0.2}$ & $0.70^{+0.10}_{-0.08}$ & $0.35^{+0.24}_{-0.11}$ \\ 
UDF2   & 53.18137 & $-$27.77758 & 2.794 & $11.1 \pm 0.2$ & $\dots$ & $ 311 \pm 28$ & $247 \pm  76$                    & $ 9.1^{+0.2}_{-0.2}$ & $0.51^{+0.12}_{-0.09}$ & $0.52^{+0.34}_{-0.16}$ \\ 
UDF3   & 53.16060 & $-$27.77628 & 2.543 & $10.3 \pm 0.2$ & X-ray & $<470      $ & $195 \pm  69$                & $ 9.1^{+0.2}_{-0.1}$ & $0.86^{+0.05}_{-0.05}$ & $0.60^{+0.37}_{-0.18}$ \\ 
UDF4   & 53.17092 & $-$27.77544 & 2.43\phantom{1}  & $10.5 \pm 0.2$ & $\dots$ & $ 116 \pm 23$ & $ 94 \pm 4$           & $ 8.6^{+0.2}_{-0.1}$ & $0.57^{+0.11}_{-0.08}$ & $0.45^{+0.27}_{-0.13}$ \\ 
UDF5   & 53.15402 & $-$27.79090 & 1.759 & $10.4 \pm 0.2$ & $\dots$ & $ 101 \pm 7\phantom{1}$ & $102 \pm 7$\phantom{1} & $ 8.7^{+0.2}_{-0.1}$ & $0.66^{+0.08}_{-0.07}$ & $0.48^{+0.24}_{-0.13}$ \\ 
UDF6   & 53.14350 & $-$27.78328 & 1.413 & $10.5 \pm 0.1$ & $\dots$ & $  80 \pm  5$ & $ \phantom{1}87 \pm  11$         & $ 8.6^{+0.2}_{-0.1}$ & $0.55^{+0.09}_{-0.07}$ & $0.44^{+0.21}_{-0.11}$ \\ 
UDF7   & 53.18052 & $-$27.77971 & 2.59\phantom{1} & $10.6 \pm 0.1$ & X-ray, Radio & $<770 $ & $\phantom{1} 56 \pm 22$ & $ 8.5^{+0.2}_{-0.2}$ & $0.44^{+0.12}_{-0.08}$ & $0.56^{+0.34}_{-0.16}$ \\ 
UDF8   & 53.16558 & $-$27.76989 & 1.546 & $11.2 \pm 0.2$ & X-ray & $< 87      $ & $149 \pm  90$                & $ 8.5^{+0.2}_{-0.1}$ & $0.17^{+0.06}_{-0.04}$ & $0.22^{+0.11}_{-0.06}$ \\ 
UDF11  & 53.16688 & $-$27.79885 & 1.998 & $10.8 \pm 0.1$ & X-ray, Radio & $<202      $ & $162 \pm  94$                & $ 8.5^{+0.2}_{-0.1}$ & $0.31^{+0.10}_{-0.06}$ & $0.18^{+0.09}_{-0.05}$ \\ 
UDF13  & 53.14616 & $-$27.77995 & 2.497 & $10.8 \pm 0.1$ & X-ray, Radio & $<166      $ & $ \phantom{1}68 \pm  18$     & $ 8.4^{+0.2}_{-0.1}$ & $0.28^{+0.10}_{-0.06}$ & $0.35^{+0.21}_{-0.10}$ \\ 
UDF16  & 53.17661 & $-$27.78551 & 1.319 & $10.9 \pm 0.1$ & $\dots$ & $  46 \pm  3$ & $ \phantom{1}40 \pm  18$         & $ 8.4^{+0.2}_{-0.1}$ & $0.24^{+0.07}_{-0.05}$ & $0.62^{+0.28}_{-0.15}$ 
\enddata
\tablecomments{IDs, stellar mass (M$_*$), SFR$_{\rm IR}$, and redshift ($z$) are from D16. Photometric redshifts are reported with two decimal points (redshifts are measured spectroscopically otherwise). The peak positions of VLA sources ($\alpha_{\rm radio}$ and $\delta_{\rm radio}$, J2000.0) are from gaussian fits (\S\ref{sec:locandsizes}). We do not tabulate the ALMA positions here because they are co-spatial with the radio positions within 40 mas (\S\ref{sec:observations}); VLA and ALMA flux densities are tabulated in Tables 2 and 3 of D16. The X-ray and radio AGN definitions are discussed in \S\ref{sec:FRC}; for AGN, SFR$_{\rm radio}$ are conservatively reported as upper limits. The dust mass M$_{\rm dust}$, gas fraction $f_{\rm gas}$, and gas-depletion time $\tau_{\rm dep}$ estimation assumes T$_{\rm dust} = 25$ K \citep{Magnelli14, Scoville16} and a gas-to-dust ratio of 100; uncertainties of these quantities reflect the range of possible values from the assumption of T$_{\rm dust} = 20 - 30$ K (details in \S\ref{sec:gas}).}
\end{deluxetable*}

\subsection{Far-Infrared/Radio Correlation at $z \sim 1-3$} \label{sec:FRC}

The far-infrared/radio correlation \citep{Helou85} has been demonstrated to hold in statistical (i.e., stacked) samples out to $z \sim 2$ \citep{Ivison10, Pannella15, Magnelli15}, but this has not been established for individual galaxies in the main-sequence of SFGs. Yet the correlation is important as a physically-motivated rationale that radio and millimeter observations trace the same extent of star-forming regions. We explore the validity of the correlation for the individual VLA-ALMA-selected SFGs by investigating their S$_{5\,{\rm cm}}$/S$_{1.3\,{\rm mm}}$ flux ratios as a function of redshift \citep[e.g.,][]{CarilliYun99} in comparison with the ratio predicted by the \citet{Rieke09} infrared SED library. To ascertain that the 5\,cm fluxes for galaxies in this test are not contaminated by AGN emission, we select a star-forming subsample by conservatively excluding all objects with L$_{\rm X}$($0.5-8$ keV) $\geqslant 3 \times 10^{42}$ erg\,s$^{-1}$ following \citet{Xue11}. That is, AGN candidates are excluded from this test on the basis of X-ray luminosity independent of whether their radio emission is enhanced over the level of star formation. 

We find that for all five galaxies in the star-forming subsample, S$_{5\,{\rm cm}}$/S$_{1.3\,{\rm mm}}$ at $z = 1-3$ is well-predicted by that of a local SED template for SFG with infrared luminosity of $10^{11.5}$ L$_{\odot}$ (Figure \ref{fig:FRC}); the rms scatter of the ratio for these five objects around this template is 0.03 dex, which suggests that the far-infrared/radio correlation does hold for these SFGs. That a single SED is a good descriptor of the flux ratio should not come as a surprise because of the small dispersion of dust temperature found in main-sequence SFGs at $z \sim 2$ \citep{Magnelli14, Scoville16}. 

On the other hand, all but one of the X-ray-selected AGN also have enhanced S$_{5\,{\rm cm}}$ over the level of S$_{1.3\,{\rm mm}}$ to varying degrees. X-ray-luminous AGN (UDF1 and 8) tend to have relatively small radio enhancement, having S$_{5\,{\rm cm}}$ similar to the levels predicted by the far-infrared/radio correlation given the S$_{1.3\,{\rm mm}}$, whereas X-ray-weak AGN candidates are among the most radio-luminous. This may be indicative of an anti-correlation between radiative and mechanical power well-established in the local Universe \citep{BestHeckman12}, although analysis of a larger radio and X-ray sample is required to test whether such a dichotomy in the Eddington ratio of AGN exists at high redshift. 

We classify as radio AGN those with S$_{5\,{\rm cm}}$ enhancement greater than two times the highest level expected from star-forming SED templates, about 0.5 dex higher than the observed S$_{5\,{\rm cm}}$/S$_{1.3\,{\rm mm}}$ ratios for the star-forming subsample, which is $\sim15\times$ the rms scatter of S$_{5\,{\rm cm}}$/S$_{1.3\,{\rm mm}}$. These radio AGN are indicated in Table 1 and in the corresponding VLA images in Figure 1. 

In the local Universe, e.g., \citet{Beck07} and \citet{Fletcher11} have shown that the morphology of SFGs is not strongly dependent on frequency at $\simeq 1.5-10$ GHz ($20 - 3$ cm). While the rest-frame $\approx 1.7$ cm continuum emission probed by our 5 cm VLA observations at $z \sim 2$ is dominated by synchrotron emission from supernovae, the thermal fraction associated with H\,\textsc{II} regions that may be present is not well constrained at these redshifts, and could be a source of morphological uncertainties. The agreement between S$_{5\,{\rm cm}}$/S$_{1.3\,{\rm mm}}$ and the template prediction suggests this is not a strong effect. Furthermore, the scale length of synchrotron and inverse-Compton losses decreases rapidly with the increasing \SFRSD\ and are $\ll 1$ kpc at the typical \SFRSD\ of $z \sim 2$ SFGs \citep{Murphy06}. Hence we expect that, for non-AGN, both VLA and ALMA should trace similar extents of star formation, i.e., if both are detected at a comparable S/N, they can be used interchangeably to measure the extinction-free star-formation sizes at $z \sim 1-3$. Although we are only able to investigate the correlation for five SFGs with SFRs $\sim 100$ \Msunyr, the success of this test highlights that it would be worth examining a larger sample of individual SFGs with future ultra-deep VLA and ALMA observations.

\begin{deluxetable*}{lcclcccc}
\tablenum{2}
\tablecaption{Size Measurement of VLA-ALMA-selected Star-Forming Galaxies \label{tab:sizemeasure}}
\tablehead{
\colhead{ID} & \colhead{S/N} & \colhead{S/N} & \colhead{VLA Deconvolved FWHM} & \colhead{ALMA Deconvolved FWHM} & \colhead{VLA r$_{1/2}$} & \colhead{ALMA r$_{1/2}$} & \colhead{Adopted r$_{\rm SF}$}\\
\colhead{} & \colhead{VLA} & \colhead{ALMA} & \colhead{$\theta_{\rm major} \times \theta_{\rm minor}$} & \colhead{$\theta_{\rm major} \times \theta_{\rm minor}$} & \colhead{(kpc)} & \colhead{(kpc)} & \colhead{(kpc)}
}
\startdata
UDF1  &	15.9	& 18.4	& $0\farcs97 \pm 0\farcs14 \times 0\farcs49 \pm 0\farcs06$ &	$0\farcs39 \pm 0\farcs04 \times 0\farcs33 \pm 0\farcs04$ &	$2.7 \pm  0.4$   & $ 	1.4 \pm  0.2 	$ &	$1.4 \pm 0.2^c $\\
UDF2  &	10.8	& 16.8	& $0\farcs42 \pm 0\farcs08 \times 0\farcs25 \pm 0\farcs04$ &	$0\farcs53 \pm 0\farcs06 \times 0\farcs45 \pm 0\farcs05$ &	$1.3 \pm  0.2$   & $	2.0 \pm  0.2	$ &	$1.6 \pm 0.2^a $\\
UDF3  &	22.9	& 14.0	& $0\farcs48 \pm 0\farcs04 \times 0\farcs27 \pm 0\farcs02$ &	$0\farcs75 \pm 0\farcs09 \times 0\farcs27 \pm 0\farcs05$ &	$1.5 \pm  0.1$   & $	1.8 \pm  0.3	$ &	$1.8 \pm 0.3^c $\\
UDF4  &	5.0	& 6.6	& $0\farcs51 \pm 0\farcs14 \phantom{.} \times < 0\farcs09$ &    $0\farcs54 \pm 0\farcs12 \times 0\farcs28 \pm 0\farcs09$ &	$2.1 \pm  0.6$   & $    1.6 \pm 0.4	$ &	$1.9 \pm 0.5^a $\\
UDF5  &	13.7	& 6.3	& $0\farcs79 \pm 0\farcs18 \times 0\farcs39 \pm 0\farcs06$ &	$0\farcs96 \pm 0\farcs25 \times 0\farcs19 \pm 0\farcs10$ &	$2.4 \pm  0.5$   & $	1.8 \pm  0.7    $ &	$2.1 \pm 0.6^a $\\
UDF6  &	16.1	& 4.9	& $0\farcs86 \pm 0\farcs12 \times 0\farcs38 \pm 0\farcs05$ &	$1\farcs06 \pm 0\farcs41 \times 0\farcs20 \pm 0\farcs17$ &	$2.4 \pm  0.3$   & $ 	2.0 \pm  1.2	$ & 	$2.2 \pm 0.7^a $\\
UDF7  &	31.4	& 4.9	& $0\farcs27 \pm 0\farcs02 \times 0\farcs22 \pm 0\farcs01$ &	$<0\farcs24  \phantom{.} \times <0\farcs13	       $ &	$1.0 \pm  0.1 $  & $	<3.0$ (2$\sigma$) &	$< 3.0^d       $\\
UDF8  &	15.3	& 4.5	& $0\farcs80 \pm 0\farcs15 \times 0\farcs29 \pm 0\farcs06$ &	$1\farcs35 \pm 0\farcs45 \times 0\farcs72 \pm 0\farcs24$ &	$2.1 \pm  0.4 $  & $	4.3 \pm  1.4 	$ &	$3.2 \pm 1.0^e $\\
UDF11 &	12.6	& 4.0	& $1\farcs08 \pm 0\farcs31 \times 0\farcs58 \pm 0\farcs16$ &	$1\farcs43 \pm 0\farcs57 \times 0\farcs69 \pm 0\farcs28$ &	$3.4 \pm  1.0 $  & $	4.2 \pm  1.7	$ &	$4.2 \pm 1.7^c $\\
      &		&       & $0\farcs39 \pm 0\farcs10 \times 0\farcs06 \pm 0\farcs03$ &                                                             &                       &   		          &	       	       \\
UDF13 &	8.8	& 3.9	& $0\farcs67 \pm 0\farcs16 \times 0\farcs40 \pm 0\farcs10$ &	$0\farcs86 \pm 0\farcs34 \times 0\farcs47 \pm 0\farcs20$ &	$2.1 \pm  0.5 $  & $	2.6 \pm  1.1	$ &     $2.6 \pm 1.1^c $\\
UDF16 &	11.9	& 3.5	& $1\farcs07 \pm 0\farcs29 \times 0\farcs60 \pm 0\farcs14$ &	$<0\farcs23     \phantom{.}      \times <0\farcs15     $ &	$3.4 \pm  0.9 $  & $ < 3.1$ (2$\sigma$)	  & 	$3.4 \pm 0.9^b $
\enddata
\tablecomments{Circularized SFG radii, r$_{\rm SF}$, are measured from the deconvolved FWHMs of VLA and/or ALMA images with the following criteria, as noted for each object in the r$_{\rm SF}$ column: (a) non-AGN, both VLA and ALMA are resolved, take average of the two bands; (b) non-AGN, only VLA is resolved, use VLA; (c) AGN, ALMA is resolved, use ALMA; (d) AGN, ALMA is unresolved, adopt 2$\sigma$ deconvolved size limit as upper limit; and (e) AGN present, but has negligible contribution to radio emission, both VLA and ALMA are resolved, same as (a). Uncertainties in the deconvolved FWHM columns are $1-\sigma$. UDF11 has two VLA components; its r$_{\rm SF}$ reflects the total size.}
\end{deluxetable*}

\begin{deluxetable}{lcccc}
\tablenum{3}
\tablecaption{Sizes and Locations of Star Formation \label{tab:sizeresult}}
\tablehead{
\colhead{ID} & \colhead{r$_{\rm SF}$} & \colhead{r$_{\rm M_*}$}   & \colhead{$\Delta_{\rm p}$(M$_*$, SF)} & \colhead{Average $\Sigma_{\rm SFR}$} \\
\colhead{}    & \colhead{(kpc)} & \colhead{(kpc)} & \colhead{(kpc)} & \colhead{(${\rm M}_\odot$yr$^{-1}$kpc$^{-2}$)}}
\startdata
UDF1  &	$1.4 \pm 0.2 $	& $1.4 \pm 0.7$	&   $0.6 \pm 0.5$  & $18.0 \pm 6.3\phantom{1}$  \\
UDF2  &	$1.6 \pm 0.2 $	& $2.5 \pm 0.9$        &   $0.7 \pm 0.5$  & $ 11.3 \pm 3.7\phantom{1}$ 	 	\\
UDF3  &	$1.8 \pm 0.3 $	& $2.7 \pm 0.5$	&   $1.7 \pm 0.5$  & $ 6.4 \pm 3.0$	 	\\
UDF4  &	$1.9 \pm 0.5 $	& $2.4 \pm 1.0$	&   $0.8 \pm 0.6$  & $ 3.3 \pm 1.8$	        \\
UDF5  &	$2.1 \pm 0.6 $	& $2.7 \pm 0.5$	&   $0.9 \pm 0.6$  & $ 2.5 \pm 1.4$	 	\\
UDF6  &	$2.2 \pm 0.7 $	& $3.5 \pm 0.7$	&   $< 0.6$        & $ 1.9 \pm 1.2$ 	 	\\
UDF7  &	$< 3.0$	        & $2.7 \pm 1.2$         &   $5.8 \pm 0.6$  & $> 2.9$		       	\\
UDF8  &	$3.2 \pm 1.0 $	& $1.8 \pm 0.4$	&   $0.8 \pm 0.6$  & $ 1.2 \pm 0.9$	 	\\
UDF11 &	$4.2 \pm 1.7 $	& $4.0 \pm 0.7$	&   $1.1 \pm 0.7$  & $ 1.0 \pm 1.0$	 	\\
UDF13 &	$2.6 \pm 1.1 $	& $1.9 \pm 0.8$	&   $0.9 \pm 0.7$  & $ 1.1 \pm 0.9$	 	\\
UDF16 &	$3.4 \pm 0.9 $	& $2.5 \pm 0.4$	&   $< 0.7$        & $ 0.4 \pm 0.2$	 	
\enddata
\tablecomments{$\Delta_{\rm p}$(M$_*$, SF) is the physical separation between the barycenters of stellar-mass and star-formation concentrations (details of sizes and separation measurements are in \S\ref{sec:locandsizes}); upper limits indicate separations smaller than the positional uncertainties. The $\Sigma_{\rm SFR}$ is averaged over the entire star-formation surface areas indicated by r$_{\rm SF}$.}
\end{deluxetable}

\subsection{Sizes and Locations of Intense Star Formation} \label{sec:locandsizes}

In Figure \ref{fig:sixpanA}, a common picture emerges that intense star formation usually occurs at the location of the stellar-mass concentration. The sizes of star-forming regions, as independently traced by the VLA and ALMA images, extend over the dominant buildup of stellar mass. In this section, we will discuss the size and location measurements to quantify these observations. None of these star-forming regions are identifiable in the rest-frame ultraviolet images, which show clumps of unobscured star formation that tend to occur in the areas peripheral to the intense obscured star formation and that comprise $0.1 - 5\%$ of the galaxy-integrated SFR (D16).

We measure the extinction-independent effective radius of the star formation distribution, r$_{\rm SF}$, from the VLA and ALMA images by means of deconvolved source sizes determined from 2D elliptical Gaussian fitting (the position, major and minor axes, flux, and position angle of the Gaussians are free parameters); multiple Gaussians are permitted, in which case all Gaussian components from the same flux `island' above 3$\sigma$ are grouped together into a source. This is carried out with PyBDSM\footnote{\url{http://www.astron.nl/citt/pybdsm}}. We also construct images at multiple $u,v$ tapering scales to accurately measure sizes of extended sources; the scale with the highest S/N for each object is adopted for the size measurement. All VLA sources except UDF11 and all ALMA sources are well described with a single component. In all cases, the typical average residual background rms values (PyBDSM's {\tt RESID\_ISL\_RMS} parameter) are $0.2\,\mu$Jy\,beam$^{-1}$ for VLA and $11\,\mu$Jy\,beam$^{-1}$ for ALMA. All VLA detections are spatially resolved and 9 of 11 of ALMA sources are resolved (all except UDF7 and 16). Comparing the circularized half-light radii, r$_{1/2}$, from VLA and ALMA (Table 2) for non-AGN, we find that except for UDF2, all of them agree within the range of uncertainties. While a larger sample is required to provide a robust comparison, this gives an early indication that radio and millimeter observations probe the same regions at $z \sim 2$. In light of this agreement, we adopt the average of r$_{1/2}$ from VLA and ALMA as the r$_{\rm SF}$ when both bands are spatially resolved and the object is non-AGN. For AGN, we adopt the ALMA r$_{1/2}$ (or the r$_{1/2}$ limit when the ALMA image is not spatially resolved) with the exception of UDF8 that harbors an X-ray AGN but does not appear to have enhanced radio emission, which we treat as non-AGN for the purpose of r$_{\rm SF}$ measurement. The VLA and ALMA r$_{1/2}$ and the band(s) used for the r$_{\rm SF}$ measurement, along with the criteria for band selection are tabulated in Table 2. 

We tested the robustness of our size measurements by injecting simulated sources into the HUDF maps. We find no bias in extracted sizes down to our S/N limit of 4, although at the lowest values of S/N ($4 - 6$) some sources are extracted with only upper limits on size. There are six ALMA sources with S/N $< 5$ (Table 2). From the simulations, we might expect to fail to measure sizes for $1 - 2$ of them, consistent with the two cases in the ALMA images (UDF7 and 16) where this has occurred. In addition, the relative uncertainties indicated by the simulations are consistent with those in Tables 2 and 3.

To compare the star-formation size with the size of stellar-mass buildup, we quantify the effective radius for the stellar-mass distribution by a circularized radius that encircles half of the stellar mass, r$_{\rm M_*}$. This is measured non-parametrically from the stellar-mass map by determining the area of the stellar-mass `island' that contains half of the total stellar mass. In the case of an ideal Gaussian, this size measure yields the same radius as Gaussian fitting, which allows us to compare the r$_{\rm M_*}$ and r$_{\rm SF}$. The r$_{\rm M_*}$ uncertainties are estimated from the rms maps from the spatially-resolved SED fitting.

The location of star-formation concentration in relation to the hosts' stellar-mass concentration is quantified by the separation between their barycenters, measured from stellar-mass and star-formation maps. The choice of band(s) used to determine the star-formation barycenter follows those of r$_{\rm SF}$ measurements, i.e., we adopt a geometric mean of the barycenters from VLA and ALMA images for non-AGN, and from the ALMA images otherwise (again, UDF8 is treated as a non-AGN for this purpose). The uncertainties are the sum in quadrature of the resolution of the stellar-mass maps ($0\farcs06$) and the uncertainties of Gaussian fitting ($\theta_{\rm beam}$/(2~SNR), see $\sigma_{\rm pos}$ discussion in \S\ref{sec:observations}).

\begin{figure*}
\figurenum{3}
\centerline{\includegraphics[width=0.88\textwidth]{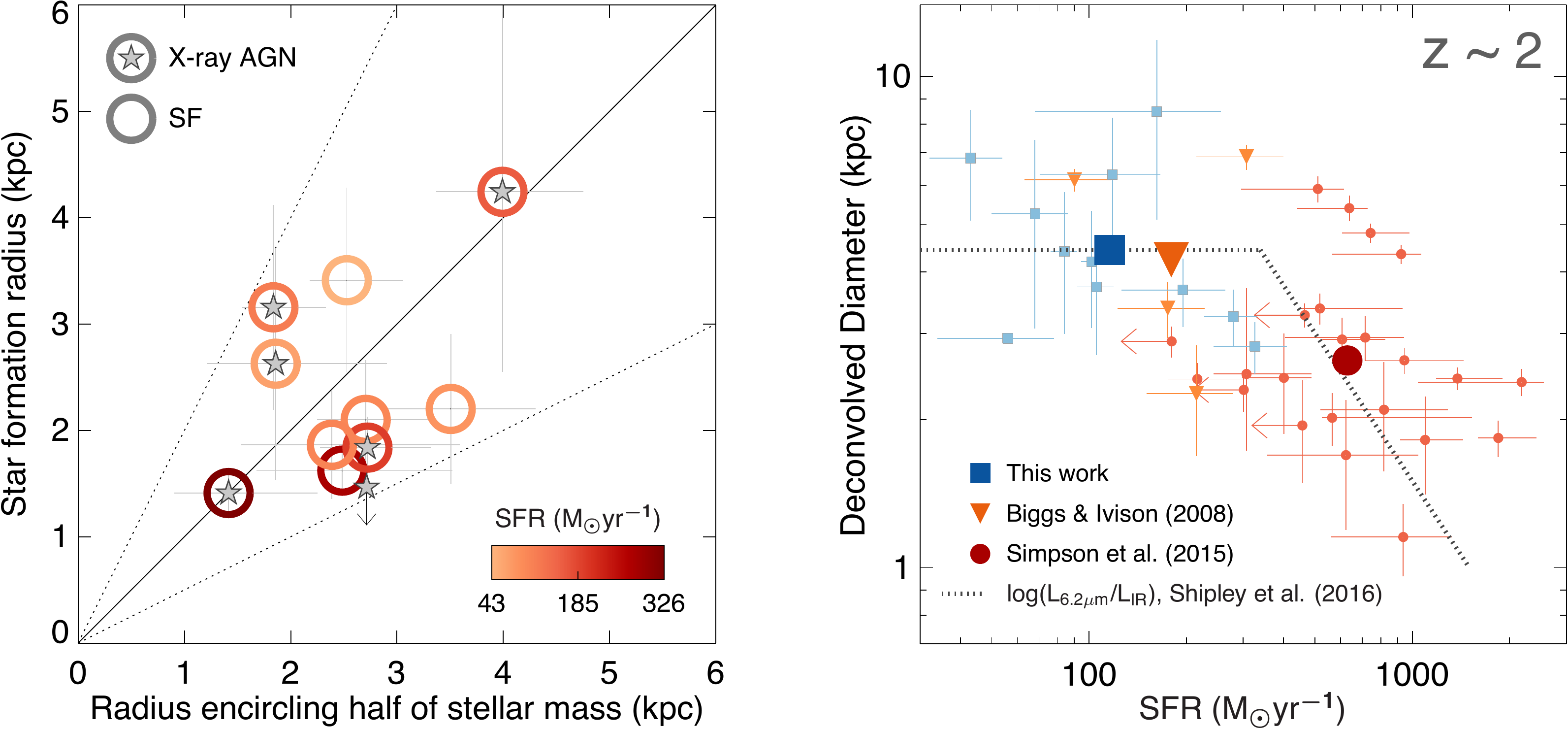}}
\caption{Left: The extinction-free sizes of SFGs in the HUDF sample broadly follow their stellar-mass size, suggesting a picture of galaxy-wide star formation, regardless of whether they harbor X-ray AGN (marked with stars). The solid and dotted lines indicate a 1:1 agreement and 2$\times$ size difference, respectively. Right: Main-sequence SFGs (this work) are $\sim2\times$ larger than submillimeter galaxies that form stars $\sim8\times$ more rapidly (the large symbols represent median for each sample). Above SFR of $200-300$ \Msunyr, SFGs becomes more compact; the dotted line shows a compactness trend inferred independently from the ratio of the 6.2 \micron\ aromatic feature's luminosity and the total infrared luminosity \citep[normalized to this work's median size]{Shipley16}.
\label{fig:SF_size}}
\end{figure*}

We find a median SFGs' r$_{\rm SF}$ of $2.1 \pm 0.9$ kpc. The separations between the star-formation and stellar-mass barycenters (Table 3) are smaller than this radius in all cases except UDF7, which appears to be interacting. On average, the star-formation and stellar-mass peaks are separated by 0.53\,r$_{\rm M_*}$ (the median separation is 0.36\,r$_{\rm M_*}$, corresponding to 0.9 kpc), indicating that star formation occurs nearly co-spatially with the stellar-mass concentration. Individually, the r$_{\rm SF}$ broadly follows the r$_{\rm M_*}$ (median r$_{\rm M_*} = 2.6 \pm 0.7$ kpc) as shown in Figure \ref{fig:SF_size} (left panel), consistent with the picture of star formation extending over a similar area as the stellar mass distribution \citep[see also,][]{Tacconi13}. Our median SFG radius is similar to those of the star-formation-dominated subsample of \citet{BiggsIvison08} SMGs, which has a median r$_{\rm SF}$ of 2.3 kpc (their star-forming subsample is reproduced in Figure \ref{fig:SF_size}). On the other hand, the SFG radii from our sample are $\approx 2-4$ times larger than those of submillimeter (sub-mm) galaxies studied with ALMA at high resolution \citep[r$_{\rm SF} = 0.7 - 1.2$ kpc,][]{Ikarashi15, Simpson15}. These sub-mm galaxies have median SFRs $3 - 8$ times those of our sample, hence an explanation is that the more compact star formation could be a characteristic of this higher SFR regime, but not for main-sequence SFGs (Figure \ref{fig:SF_size}). 

The increasing compactness of SFGs at SFR $\gtrsim 300$ \Msunyr\ is independently indicated by the decreasing ratio of mid-infrared aromatic luminosity to the total infrared luminosity \citep[L$_{6.2 \micron}$/L$_{\rm IR}$ trend shown in Figure \ref{fig:SF_size};][]{Nordon12, Pope13, Shipley16}. Consistent with this trend, the two most intense SFGs in our sample (UDF1 and 2) are also the most compact. While the origin of such a trend is at present unclear, if the star-formation rate in the interstellar medium is regulated by some internal process (such as a form of stellar feedback), then this result may be expected.  At the centers of massive starbursts, the surface densities rise and the vertical weight on a cloud can hinder the effectiveness of internal feedback processes at dispersing gas \citep[e.g.,][]{OstrikerShetty11, Hopkins13}. In these environments, the star-formation efficiency can rise dramatically, decreasing the effective radius for star formation within the galaxy.

A major result from our observations is that main sequence SFGs at $z \sim 2$ exhibit vigorous star formation in a distributed manner over size scales of a few kiloparsec, which is to be contrasted with the typical sub-kpc size scales of comparably luminous galaxies in the local Universe \citep{Condon91, Rujo11, Lutz16}. Such a result is expected from numerical simulations of galaxy growth via in-situ star formation that is fed by the accretion of gas from the intergalactic medium \citep[e.g.,][]{Dekel09, Dave10}. High-resolution cosmological simulations have found that star formation may remain distributed even in the most extreme systems at this epoch in the submillimeter-selected population \citep{Narayanan15}.

The similarity of the effective radii of star-formation and stellar-mass in our field-galaxy sample is to be contrasted with recent ALMA high-resolution studies that target high-SFR galaxies. \citet{Barro16} found that the 870 \micron\ sizes of main-sequence SFGs (median SFR $\sim 300$ \Msunyr) selected by their compact optical sizes to represent the progenitors of compact galaxies at $z\sim2$ have r$_{\rm SF} \approx$ 1 kpc, and do not show the star-formation size trend with stellar-mass distribution sizes. Similarly, the \citet{Tadaki16} 870 \micron\ observations of H$\alpha$-selected SFGs at $z = 2.2-2.5$ (median SFR $\sim 230$ \Msunyr, also in the main-sequence) found their star-formation radii to be $< 1.5$ kpc, a factor of two smaller than their average rest-frame optical size (3.2 kpc). These two samples exhibit a similar characteristic of star formation distribution to our sample in that their star-formation concentrations also peak at the same locations as the hosts' stellar-mass concentrations, but they appear to lack the extended component of galaxy-wide star formation. A possible explanation is that the extended star formation found in field galaxies could be a distinct evolutionary state from (and possibly preceding) the compact star formation in the \citet{Barro16} and \citet{Tadaki16} samples, with the latter representing a state closer to the conclusion of bulge formation and the subsequent cessation of the bulk of star formation \citep[e.g.,][]{Barro15}.

\subsection{Dust and Gas Properties} \label{sec:gas}

We estimate the dust masses from the 1.3 mm fluxes using the \citet{LiDraine01} dust mass absorption coefficients. The dominant source of uncertainty in dust mass estimates is the dust temperature T$_{\rm dust}$, which is not known precisely for our sample, but {\it Herschel} studies have shown that T$_{\rm dust}$ of main-sequence SFGs at $z \sim 2$ is in the range of $20 - 30$ K \citep{Dunne11, Auld13}, prompting us to adopt T$_{\rm dust} = 25$ K \citep{Magnelli14, Kirkpatrick15, Scoville16} and estimate the uncertainties using T$_{\rm dust}  =$ 20 and 30 K for upper and lower limits, respectively. Since the ALMA measurements are in the Rayleigh-Jeans regime, the dust masses have only a moderate dependence on the assumed temperature \citep{Hildebrand83, Casey12, Elia16}. Dust masses are in the range of log(M$_{\rm dust}$/\Msun) $= 8.4 - 9.1$ with a typical uncertainty of 0.2 dex (tabulated in Table \ref{tab:sourcetable}). The dust mass estimates are in agreement with the \citet{Magdis12} results of stacking analysis for $z \sim 2$ main-sequence SFGs, suggesting that the dust masses of our HUDF sample are representative of main-sequence populations at $z \sim 2$.

If we assume a gas-to-dust ratio of 100 \citep{Leroy11, Magdis12}, the gas fraction, $f_{\rm gas} = M_{\rm gas}/(M_{\rm star} + M_{\rm gas})$, of our sample is on average $0.5 \pm 0.2$, indicating the gas-rich nature of these systems similar to the normal SFGs at $z \sim 2$ studied by \citet{Tacconi13}, though significantly more gas rich than typical sub-mm galaxies \citep[e.g.,][]{Narayanan12}. The average gas depletion time, $\tau_{\rm dep} = M_{\rm gas}$/SFR, of our sample is $0.4 \pm 0.2$ Gyr, comparable to their average mass-doubling time of 0.4 Gyr, assuming that star formation-driven outflows do not result in irreversible mass losses. These $\tau_{\rm dep}$ are similar to the main-sequence samples of \citet{Sargent14} and \citet{Silverman15}, further assuring the main-sequence nature of our sample independent of the presence of AGN. We note that if the T$_{\rm dust}$ is warmer than the assumed 25 K, the gas fraction could be lower (e.g., T$_{\rm dust} = 30$ K implies average $f_{\rm gas}$ and $\tau_{\rm dep}$ of $0.4 \pm 0.2$ and $0.3 \pm 0.1$ Gyr) and hence requires a significant amount of cold gas accretion to sustain the current level of star formation for a significant period of time. However, the metallicities of our SFGs are expected to be lower than for local analogs \citep[e.g.,][]{Zahid13}. The rapid reduction in dust mass with decreasing metallicity \citep{Draine07} would then argue for larger gas fractions in these galaxies.

The co-spatiality and the similar effective radii of radio and millimeter emissions imply that in most cases, the cold gas associated to star formation resides in the star-forming regions, which we have shown occur near the stellar-mass concentrations. Hence, if this level of SFR is sustained through a mass-doubling time, we expect that the newly formed stellar masses will occur near the existing stellar-mass concentrations, i.e., enhancing the stellar mass surface density near the central region, consistent with a picture of in-situ bulge formation \citep[see also,][]{Barro16, Tadaki16}.

\subsection{\SFRSD\ Implies Galaxy-Wide Star Formation-Driven Outflows?} \label{sec:sigmasfr}

The spatially-resolved \SFRSD\ for 9 of 11 galaxies peaks within $\leqslant 1.1$ kpc of the stellar-mass distribution (with the exception of UDF3 and 7; Table 3), and all have \SFRSD\ in the range of $0.4 - 18$ \Msunyrkpcsq\ (median \SFRSD\ $= 2.5$ \Msunyrkpcsq), the higher end of which is comparable to local starbursts \citep{Kennicutt98}. The level of \SFRSD\ found in our main-sequence sample is significantly below those found in luminous sub-mm galaxies of $90-100$ \Msunyrkpcsq\ \citep{Ikarashi15, Simpson15}, suggesting very different conditions for star formation between them.

Recent studies show that \SFRSD\ is a strong predictor of star formation-driven outflows at $z = 1-2$. An increased broad-fraction of H$\alpha$ line flux at $z \sim 2$ associated with star formation-driven outflows is found to strongly depend on \SFRSD, with a \SFRSD\ of 1 \Msunyrkpcsq\ being the threshold above which the large broad line fractions are observed \citep{Newman12}. At a similar \SFRSD\ threshold of $\approx 0.3$ \Msunyrkpcsq, \citet{Bordoloi14} also found a significant increase in the equivalent width of the Mg II absorption line at $z \sim 1$ that indicates cool outflowing gas. If \SFRSD\ is indeed a predictor of star formation-driven outflows at these redshifts, we expect that main-sequence SFGs like those in our sample, whose galaxy-average \SFRSD\ are larger than these threshold levels (Table 3), will harbor outflows across a significant extent comparable to the areas of stellar-mass buildup (i.e., galaxy-wide). While the {\it presence} of outflows appears to depend on \SFRSD, how does outflow {\it rate} depend on \SFRSD\ is not well established, especially for molecular gas in a turbulent galactic environment at $z \sim 2$. If a strong relationship also exists between them, we expect that the areas exhibiting large \SFRSD\ near galaxy centers will also be the areas with the largest outflows. Future spatially-resolved observations of cold gas dynamics could provide definitive evidence whether this intense, galaxy-wide star formation also drives galaxy-wide outflows.

\subsection{Remarks on star formation in AGN hosts at $z \sim 2$} \label{sec:SFinAGN}

In \S\ref{sec:FRC}, we have shown that the 4 Ms {\it Chandra} observations in the HUDF allows us to detect X-ray AGN candidates residing in main-sequence SFGs with varying degrees of radio emission enhancement over the level corresponding to their SFRs. Six of 11 galaxies in our SFG sample are detected in X-rays, with the most luminous reaching L$_{\rm 0.5-8\,keV} \simeq 5\times10^{43}$ ergs\,s$^{-1}$. Two among the X-ray detected SFGs (UDF3 and 11) are only detected in the soft X-ray band ($0.5-2$ keV), suggesting that their X-ray emission could originate solely from star formation. However, UDF11 also has radio emission enhanced by more than twice the level associated with star formation (Figure 2 Left; this level of radio enhancement is $15\times$ the scatter around the emission level implied by the far-IR/radio correlation, \S\ref{sec:FRC}) that suggests a presence of radio AGN.

Omitting the only ambiguous case of UDF3 from the AGN sub-sample, the AGN fraction (5 of 11 galaxies) is still higher than typically reported \citep[e.g.,][]{Xue10} at these redshifts, but not unexpected since AGN activity is known to depend on both stellar mass and gas mass as evident by its clear dependence on star formation of its host even on a galaxy-wide scale \citep[e.g.,][]{Silverman09, Brusa09}. With the ALMA continuum detections at 1.3 mm preferential to the most massive galaxies with high SFRs at $z \sim 1 - 3$ (D16), such selection is also effective at probing rapidly-growing supermassive black holes at $z \sim 2$ including those with mild obscuration. Further support of such a high detection rate of AGN using ALMA comes from \citet{Umehata15} who report a high X-ray AGN fraction in the SSA22 field.

Despite the high sensitivity of the {\it Chandra} observations in the HUDF, the X-ray counts in the full $0.5-8$ keV band for 4 of 6 AGN candidates are still low ($16 - 26$ counts), highlighting the necessity of sensitive X-ray observations to characterize AGN populations in main-sequence SFGs at $z \sim 2$. In the remaining two sources with sufficient count statistics, UDF1 and UDF 8, which have 434 and 878 full-band counts, respectively, the column densities are $1.5\times10^{21}$ and $7\times10^{22}$ cm$^{-2}$ respectively, based on {\tt Web-PIMMS}\footnote{\url{http://heasarc.gsfc.nasa.gov/Tools}}, provided by HEASARC, assuming an intrinsic photon index of 2 and the redshift of the source, indicating mild obscuration. 

Independent of the AGN selection method, ALMA imaging shows that all AGN candidates in the sample contain large amounts of cold dust and gas (Table 1) associated with star formation. The trend of the star formation size with SFR shown in Figure 3 is consistent with the ALMA 870 \micron\ sizes of star formation in the hosts of six X-ray detected AGN that have SFRs of $\approx130 - 400$ \Msunyr\ \citep{Harrison16}. 

For UDF7, 11, and 13 where the radio flux enhancement allows pinpointing the AGN location using the VLA images, we find that the AGN are within 0.5 kpc of the barycenter of star formation of their host galaxies as measured from the ALMA images (an upper limit of given by our ability to pinpoint AGN and star formation given their S/N). For UDF11 and 13, the star-formation concentrations are $\approx1$ kpc from the stellar-mass barycenters, whereas UDF7 is notable for these activities being $\approx6$ kpc away from the dominant stellar-mass concentration, a possible consequence of galaxy interactions.

\section{Conclusion} \label{conclude}

We have made the first comparison between ultra-deep VLA and ALMA imaging of $z \sim 2$ main-sequence SFGs. The far-infrared/radio correlation appears to hold for individual main-sequence SFGs with SFR $\sim$ 100 \Msunyr\ at $z \sim 2$ and the extinction-independent distributions of star formation are consistent between the data sets. 

The intense star formation in our blank-field-selected sample extends over a large galactic area regardless of their stellar-mass morphology (isolated or morphologically disturbed) or the presence of AGN, with the SFR and dust-mass surface densities both peaking near the existing stellar-mass concentration. These findings provide direct-imaging evidence of a gas-rich galactic environment with widespread occurrence of intense star formation. The spatially-resolved SFR surface densities are sufficiently large across the areas of dominant stellar-mass buildups that they may drive galaxy-wide outflows. Where radio excess permits pinpointing of the AGN, it is found to be co-spatial with the dust mass concentrations.

The median star-formation diameter in our main-sequence SFGs sample is $4.2 \pm 1.8$ kpc, two times larger than those of sub-mm galaxies forming stars at $3-8$ times higher than the main-sequence SFGs, indicating that, at $z \sim 2$, the SFR threshold above which a significant population of more compact SFGs appears to emerge is $\sim300$ \Msunyr.

\section*{Acknowledgments}
We would like to thank the anonymous referee for comments and suggestions to improve the paper, as well as E. Murphy for helpful discussions. This work was supported by the World Premier International Research Center Initiative (WPI), MEXT, Japan and JSPS KAKENHI Grant Number JP15K17604. WR acknowledges support from Chulalongkorn University's Ratchadapiseksompot Endowment Fund and CUniverse (CUAASC). JSD acknowledges the support of the European Research Council via the award of an Advanced Grant (PI J. Dunlop), and the contribution of the EC FP7 SPACE project ASTRODEEP (Ref. No: 312725). RJI acknowledges support from the European Research Council through the Advanced Grant COSMICISM 321302. RJM acknowledges ERC funding via the award of a consolidator grant (PI McLure). AK gratefully acknowledges support from the YCAA Prize Postdoctoral Fellowship. Partial support for DN was provided by NSF AST-1442650, NASA HST AR-13906.001 and a Cottrell College Science Award. The National Radio Astronomy Observatory is a facility of the National Science Foundation operated under cooperative agreement by Associated Universities, Inc; this paper makes use of VLA data from program VLA/14A-360. This paper makes use of the following ALMA data: ADS/JAO.ALMA\#2012.1.00173.S. ALMA is a partnership of ESO (representing its member states), NSF (USA) and NINS (Japan), together with NRC (Canada) and NSC and ASIAA (Taiwan) and KASI (Republic of Korea), in cooperation with the Republic of Chile. The Joint ALMA Observatory is operated by ESO, AUI/NRAO and NAOJ. This work is based in part on observations made with the NASA/ESA {\it Hubble Space Telescope}, which is operated by the Association of Universities for Research in Astronomy, Inc, under NASA contract NAS5-26555.




\newpage
\begin{thebibliography}{}
\bibitem[Auld et al.(2013)]{Auld13} Auld, R., Bianchi, S., Smith, M.~W.~L., et al.\ 2013, \mnras, 428, 1880 
\bibitem[Barro et al.(2015)]{Barro15} Barro, G., Faber, S.~M., Koo, D.~C., et al.\ 2015, arXiv:1509.00469 
\bibitem[Barro et al.(2016)]{Barro16} Barro, G., Kriek, M., P{\'e}rez-Gonz{\'a}lez, P.~G., et al.\ 2016, arXiv:1607.01011 
\bibitem[Beck(2007)]{Beck07} Beck, R.\ 2007, \aap, 470, 539 
\bibitem[Bell(2003)]{Bell03} Bell, E.~F.\ 2003, \apj, 586, 794
\bibitem[Best \& Heckman(2012)]{BestHeckman12} Best, P.~N., \& Heckman, T.~M.\ 2012, \mnras, 421, 1569 
\bibitem[Bordoloi et al.(2014)]{Bordoloi14} Bordoloi, R., Lilly, S.~J., Hardmeier, E., et al.\ 2014, \apj, 794, 130
\bibitem[Biggs \& Ivison(2008)]{BiggsIvison08} Biggs, A.~D., \& Ivison, R.~J.\ 2008, \mnras, 385, 893 
\bibitem[Bournaud \& Elmegreen(2009)]{Bournaud09} Bournaud, F., \& Elmegreen, B.~G.\ 2009, \apjl, 694, L158
\bibitem[Bournaud et al.(2014)]{Bournaud14} Bournaud, F., Perret, V., Renaud, F., et al.\ 2014, \apj, 780, 57
\bibitem[Brusa et al.(2009)]{Brusa09} Brusa, M., Fiore, F., Santini, P., et al.\ 2009, \aap, 507, 1277 
\bibitem[Carilli \& Yun(1999)]{CarilliYun99} Carilli, C.~L., \& Yun, M.~S.\ 1999, \apjl, 513, L13 
\bibitem[Casey(2012)]{Casey12} Casey, C.~M.\ 2012, \mnras, 425, 3094 
\bibitem[Chabrier(2003)]{Chabrier03} Chabrier, G.\ 2003, \pasp, 115, 763 
\bibitem[Cibinel et al.(2015)]{Cibinel15} Cibinel, A., Le Floc'h, E., Perret, V., et al.\ 2015, \apj, 805, 181 
\bibitem[Condon et al.(1991)]{Condon91} Condon, J.~J., Huang, Z.-P., Yin, Q.~F., \& Thuan, T.~X.\ 1991, \apj, 378, 65 
\bibitem[Condon(1997)]{Condon97} Condon, J.~J.\ 1997, \pasp, 109, 166 
\bibitem[Conselice(2003)]{Conselice03} Conselice, C.~J.\ 2003, \apjs, 147, 1 
\bibitem[Dav{\'e} et al.(2010)]{Dave10} Dav{\'e}, R., Finlator, K., Oppenheimer, B.~D., et al.\ 2010, \mnras, 404, 1355 
\bibitem[Dekel et al.(2009)]{Dekel09} Dekel, A., Birnboim, Y., Engel, G., et al.\ 2009, \nat, 457, 451
\bibitem[Draine et al.(2007)]{Draine07} Draine, B.~T., Dale, D.~A., Bendo, G., et al.\ 2007, \apj, 663, 866 
\bibitem[Dunlop et al.(2016)]{Dunlop16} Dunlop, J.~S., McLure, R.~J., Biggs, A.~D., et al.\ 2016, arXiv:1606.00227 
\bibitem[Dunne et al.(2011)]{Dunne11} Dunne, L., Gomez, H.~L., da Cunha, E., et al.\ 2011, \mnras, 417, 1510 
\bibitem[Elbaz et al.(2011)]{Elbaz11} Elbaz, D., Dickinson, M., Hwang, H.~S., et al.\ 2011, \aap, 533, A119 
\bibitem[Elia \& Pezzuto(2016)]{Elia16} Elia, D., \& Pezzuto, S.\ 2016, \mnras, 461, 1328 
\bibitem[Ellis et al.(2013)]{Ellis13} Ellis, R.~S., McLure, R.~J., Dunlop, J.~S., et al.\ 2013, \apjl, 763, L7 
\bibitem[Elvis et al.(1994)]{Elvis94} Elvis, M., Wilkes, B.~J., McDowell, J.~C., et al.\ 1994, \apjs, 95, 1 
\bibitem[F{\"o}rster Schreiber et al.(2009)]{FS09} F{\"o}rster Schreiber, N.~M., Genzel, R., Bouch{\'e}, N., et al.\ 2009, \apj, 706, 1364 
\bibitem[Fletcher et al.(2011)]{Fletcher11} Fletcher, A., Beck, R., Shukurov, A., Berkhuijsen, E.~M., \& Horellou, C.\ 2011, \mnras, 412, 2396 
\bibitem[Genel et al.(2012)]{Genel12} Genel, S., Naab, T., Genzel, R., et al.\ 2012, \apj, 745, 11 
\bibitem[Grogin et al.(2011)]{Grogin11} Grogin, N.~A., Kocevski, D.~D., Faber, S.~M., et al.\ 2011, \apjs, 197, 35 
\bibitem[Guo et al.(2012)]{Guo12} Guo, Y., Giavalisco, M., Ferguson, H.~C., Cassata, P., \& Koekemoer, A.~M.\ 2012, \apj, 757, 120 
\bibitem[Guo et al.(2015)]{Guo15} Guo, Y., Ferguson, H.~C., Bell, E.~F., et al.\ 2015, \apj, 800, 39 
\bibitem[Harrison et al.(2016)]{Harrison16} Harrison, C.~M., Simpson, J.~M., Stanley, F., et al.\ 2016, \mnras, 457, L122 
\bibitem[Hayward et al.(2013)]{Hayward13} Hayward, C.~C., Narayanan, D., Kere{\v s}, D., et al.\ 2013, \mnras, 428, 2529 
\bibitem[Helou et al.(1985)]{Helou85} Helou, G., Soifer, B.~T., \& Rowan-Robinson, M.\ 1985, \apjl, 298, L7 
\bibitem[Hildebrand(1983)]{Hildebrand83} Hildebrand, R.~H.\ 1983, \qjras, 24, 267 
\bibitem[Hopkins et al.(2010)]{Hopkins10} Hopkins, P.~F., Younger, J.~D., Hayward, C.~C., Narayanan, D., \& Hernquist, L.\ 2010, \mnras, 402, 1693 
\bibitem[Hopkins et al.(2013)]{Hopkins13} Hopkins, P.~F., Cox, T.~J., Hernquist, L., et al.\ 2013, \mnras, 430, 1901 
\bibitem[Ikarashi et al.(2015)]{Ikarashi15} Ikarashi, S., Ivison, R.~J., Caputi, K.~I., et al.\ 2015, \apj, 810, 133 
\bibitem[Ivison et al.(2010)]{Ivison10} Ivison, R.~J., Magnelli, B., Ibar, E., et al.\ 2010, \aap, 518, L31
\bibitem[Kartaltepe et al.(2012)]{Kartaltepe12} Kartaltepe, J.~S., Dickinson, M., Alexander, D.~M., et al.\ 2012, \apj, 757, 23
\bibitem[Kirkpatrick et al.(2015)]{Kirkpatrick15} Kirkpatrick, A., Pope, A., Sajina, A., et al.\ 2015, \apj, 814, 9 
\bibitem[Kennicutt(1998)]{Kennicutt98} Kennicutt, R.~C., Jr.\ 1998, \apj, 498, 541
\bibitem[Kere{\v s} et al.(2005)]{Keres05} Kere{\v s}, D., Katz, N., Weinberg, D.~H., \& Dav{\'e}, R.\ 2005, \mnras, 363, 2
\bibitem[Koekemoer et al.(2011)]{Koekemoer11} Koekemoer, A.~M., Faber, S.~M., Ferguson, H.~C., et al.\ 2011, \apjs, 197, 36
\bibitem[Koekemoer et al.(2013)]{Koekemoer13} Koekemoer, A.~M., Ellis, R.~S., McLure, R.~J., et al.\ 2013, \apjs, 209, 3 
\bibitem[Leroy et al.(2011)]{Leroy11} Leroy, A.~K., Bolatto, A., Gordon, K., et al.\ 2011, \apj, 737, 12 
\bibitem[Li \& Draine(2001)]{LiDraine01} Li, A., \& Draine, B.~T.\ 2001, \apj, 554, 778  
\bibitem[Lotz et al.(2004)]{Lotz04} Lotz, J.~M., Primack, J., \& Madau, P.\ 2004, \aj, 128, 163
\bibitem[Lutz et al.(2016)]{Lutz16} Lutz, D., Berta, S., Contursi, A., et al.\ 2016, \aap, 591, A136 
\bibitem[Magdis et al.(2012)]{Magdis12} Magdis, G.~E., Daddi, E., B{\'e}thermin, M., et al.\ 2012, \apj, 760, 6 
\bibitem[Magnelli et al.(2014)]{Magnelli14} Magnelli, B., Lutz, D., Saintonge, A., et al.\ 2014, \aap, 561, A86 
\bibitem[Magnelli et al.(2015)]{Magnelli15} Magnelli, B., Ivison, R.~J., Lutz, D., et al.\ 2015, \aap, 573, A45 
\bibitem[McMullin et al.(2007)]{McMullin07} McMullin, J.~P., Waters, B., Schiebel, D., Young, W., \& Golap, K.\ 2007, Astronomical Data Analysis Software and Systems XVI, 376, 127 
\bibitem[Mor \& Netzer(2012)]{MorNetzer12} Mor, R., \& Netzer, H.\ 2012, \mnras, 420, 526 
\bibitem[Mullaney et al.(2011)]{Mullaney11} Mullaney, J.~R., Alexander, D.~M., Goulding, A.~D., \& Hickox, R.~C.\ 2011, \mnras, 414, 1082 
\bibitem[Murphy et al.(2006)]{Murphy06} Murphy, E.~J., Helou, G., Braun, R., et al.\ 2006, \apjl, 651, L111 
\bibitem[Narayanan et al.(2012)]{Narayanan12} Narayanan, D., Bothwell, M., \& Dav{\'e}, R.\ 2012, \mnras, 426, 1178 
\bibitem[Narayanan et al.(2015)]{Narayanan15} Narayanan, D., Turk, M., Feldmann, R., et al.\ 2015, \nat, 525, 496 
\bibitem[Nelson et al.(2015)]{Nelson15} Nelson, E.~J., van Dokkum, P.~G., F{\"o}rster Schreiber, N.~M., et al.\ 2015, arXiv:1507.03999 
\bibitem[Nelson et al.(2016)]{Nelson16} Nelson, E.~J., van Dokkum, P.~G., Momcheva, I.~G., et al.\ 2016, \apjl, 817, L9 
\bibitem[Newman et al.(2012)]{Newman12} Newman, S.~F., Genzel, R., F{\"o}rster-Schreiber, N.~M., et al.\ 2012, \apj, 761, 43
\bibitem[Noeske et al.(2007)]{Noeske07} Noeske, K.~G., Weiner, B.~J., Faber, S.~M., et al.\ 2007, \apjl, 660, L43 
\bibitem[Nordon et al.(2012)]{Nordon12} Nordon, R., Lutz, D., Genzel, R., et al.\ 2012, \apj, 745, 182 
\bibitem[Ostriker \& Shetty(2011)]{OstrikerShetty11} Ostriker, E.~C., \& Shetty, R.\ 2011, \apj, 731, 41 
\bibitem[Pannella et al.(2015)]{Pannella15} Pannella, M., Elbaz, D., Daddi, E., et al.\ 2015, \apj, 807, 141 
\bibitem[Pope et al.(2013)]{Pope13} Pope, A., Wagg, J., Frayer, D., et al.\ 2013, \apj, 772, 92 
\bibitem[Reddy et al.(2010)]{Reddy10} Reddy, N.~A., Erb, D.~K., Pettini, M., Steidel, C.~C., \& Shapley, A.~E.\ 2010, \apj, 712, 1070 
\bibitem[Rieke et al.(2009)]{Rieke09} Rieke, G.~H., Alonso-Herrero, A., Weiner, B.~J., et al.\ 2009, \apj, 692, 556 
\bibitem[Rodighiero et al.(2011)]{Rodighiero11} Rodighiero, G., Daddi, E., Baronchelli, I., et al.\ 2011, \apjl, 739, L40 
\bibitem[Rujopakarn et al.(2011)]{Rujo11} Rujopakarn, W., Rieke, G.~H., Eisenstein, D.~J., \& Juneau, S.\ 2011, \apj, 726, 93 
\bibitem[Sargent et al.(2014)]{Sargent14} Sargent, M.~T., Daddi, E., B{\'e}thermin, M., et al.\ 2014, \apj, 793, 19 
\bibitem[Scoville et al.(2016)]{Scoville16} Scoville, N., Sheth, K., Aussel, H., et al.\ 2016, \apj, 820, 83 
\bibitem[Silverman et al.(2009)]{Silverman09} Silverman, J.~D., Lamareille, F., Maier, C., et al.\ 2009, \apj, 696, 396 
\bibitem[Silverman et al.(2015)]{Silverman15} Silverman, J.~D., Daddi, E., Rodighiero, G., et al.\ 2015, \apjl, 812, L23 
\bibitem[Shipley et al.(2016)]{Shipley16} Shipley, H.~V., Papovich, C., Rieke, G.~H., Brown, M.~J.~I., \& Moustakas, J.\ 2016, \apj, 818, 60 
\bibitem[Simpson et al.(2015)]{Simpson15} Simpson, J.~M., Smail, I., Swinbank, A.~M., et al.\ 2015, \apj, 799, 81 
\bibitem[Speagle et al.(2014)]{Speagle14} Speagle, J.~S., Steinhardt, C.~L., Capak, P.~L., \& Silverman, J.~D.\ 2014, \apjs, 214, 15 
\bibitem[Stott et al.(2016)]{Stott16} Stott, J.~P., Swinbank, A.~M., Johnson, H.~L., et al.\ 2016, \mnras, 457, 1888 
\bibitem[Tacconi et al.(2013)]{Tacconi13} Tacconi, L.~J., Neri, R., Genzel, R., et al.\ 2013, \apj, 768, 74 
\bibitem[Tadaki et al.(2016)]{Tadaki16} Tadaki, K.-i., Genzel, R., Kodama, T., et al.\ 2016, arXiv:1608.05412 
\bibitem[Umehata et al.(2015)]{Umehata15} Umehata, H., Tamura, Y., Kohno, K., et al.\ 2015, \apjl, 815, L8 
\bibitem[Whitaker et al.(2012)]{Whitaker12} Whitaker, K.~E., van Dokkum, P.~G., Brammer, G., \& Franx, M.\ 2012, \apjl, 754, L29 
\bibitem[Wisnioski et al.(2015)]{Wisnioski15} Wisnioski, E., F{\"o}rster Schreiber, N.~M., Wuyts, S., et al.\ 2015, \apj, 799, 209 
\bibitem[Wuyts et al.(2012)]{Wuyts12} Wuyts, S., F{\"o}rster Schreiber, N.~M., Genzel, R., et al.\ 2012, \apj, 753, 114
\bibitem[Xue et al.(2010)]{Xue10} Xue, Y.~Q., Brandt, W.~N., Luo, B., et al.\ 2010, \apj, 720, 368 
\bibitem[Xue et al.(2011)]{Xue11} Xue, Y.~Q., Luo, B., Brandt, W.~N., et al.\ 2011, \apjs, 195, 10 
\bibitem[Zahid et al.(2013)]{Zahid13} Zahid, H.~J., Geller, M.~J., Kewley, L.~J., et al.\ 2013, \apjl, 771, L19 
\bibitem[Zamojski et al.(2007)]{Zamojski07} Zamojski, M.~A., Schiminovich, D., Rich, R.~M., et al.\ 2007, \apjs, 172, 468 
\end{thebibliography}
\end{document}